\documentclass[conference]{IEEEtran}

\usepackage{color}
\usepackage{graphicx}
\usepackage{epstopdf}
\usepackage{amsmath}
\usepackage{amssymb}
\usepackage{hyperref}
\usepackage[english]{babel}
\usepackage{cite}
\usepackage{rotfloat}
\usepackage{mathtools}
\usepackage{amsmath}
\usepackage{makecell}
\usepackage{algorithm,algorithmic}
\usepackage{multirow}
\usepackage{subfigure}
\usepackage{booktabs}
\usepackage{colortbl}
\usepackage{multirow}
\usepackage{hhline}
\usepackage{stfloats}
\usepackage{multicol}
\usepackage[justification=centering,font=small]{caption}

\newsavebox{\foobox}

\newcommand{\setA}{\mathbb{A}}

\definecolor{kugray5}{RGB}{224,224,224}

\usepackage[normalem]{ulem}
\newcommand\rsout{\bgroup\markoverwith
	{\textcolor{red}{\rule[0.5ex]{2pt}{0.8pt}}}\ULon}



\makeatletter
\newcommand{\ALOOP}[1]{\ALC@it\algorithmicloop\ #1%
	\begin{ALC@loop}}
	\newcommand{\ENDALOOP}{\end{ALC@loop}\ALC@it\algorithmicendloop}

\makeatother

\usepackage{etoolbox}
\let\mybibitem\bibitem
\renewcommand{\bibitem}[1]{%
	\ifstrequal{#1}{nature}
	{\color{blue}\mybibitem{#1}}
	{\color{black}\mybibitem{#1}}%
}

\graphicspath{ {Figures/} }

\newtheorem{remark}{Remark}

\DeclareCaptionLabelSeparator{periodspace}{.\quad}

\addto\captionsenglish{}
\newcommand\nbthis{\addtocounter{equation}{1}\tag{\theequation}}

\newcommand{\norm}[1]{\left\lVert#1\right\rVert} 
\newcommand{\abs}[1]{\left|#1\right|} 

\newcommand{\tr}[1]{\mathrm{trace}\left(#1\right)} 
\newcommand{\diag}[1]{\mathrm{diag}\left\{#1\right\}} 

\newcommand{\re}[1]{\mathfrak{R}{\left(#1\right)}}

\newcommand{\mean}[1]{\mathbb{E} \left\{#1\right\}}

\newcommand{\prob}{\left(\mathrm{P} \right)}
\newcommand{\probn}{\left(\mathrm{Pn} \right)}

\newcommand{\mQ}{\textbf{\textit{Q}}}
\newcommand{\mR}{\textbf{\textit{R}}}
\newcommand{\mH}{\textbf{\textit{H}}} 

\newcommand{\mA}{\textbf{\textit{A}}}

\newcommand{\mI}{\textbf{\textit{I}}}

\newcommand{\mB}{\textbf{\textit{B}}}
\newcommand{\mC}{\textbf{\textit{C}}}
\newcommand{\mD}{\textbf{\textit{D}}}

\newcommand{\mU}{\textbf{\textit{U}}}
\newcommand{\mV}{\textbf{\textit{V}}}
\newcommand{\mE}{\textbf{\textit{E}}}
\newcommand{\mHr}{\textbf{\textit{H}}_r} 
\newcommand{\mHt}{\textbf{\textit{H}}_t} 
\newcommand{\mIr}{\textbf{\textit{I}}_{N_r}}

\newcommand{\setC}{\mathbb{C}} 

\newcommand{\vx}{\textbf{\textit{x}}}
\newcommand{\vy}{\textbf{\textit{y}}}
\newcommand{\vr}{\textbf{\textit{r}}}

\newcommand{\vn}{\textbf{\textit{n}}}

\newcommand{\vt}{\textbf{\textit{t}}}

\newcommand{\bPhi}{\boldsymbol{\Phi}}
\newcommand{\bUpsilon}{\boldsymbol{\Upsilon}}

\newcommand{\bPsi}{\boldsymbol{\Psi}}

\newcommand{\Pa}{P_{\mathrm{a}}} 
\newcommand{\Pamax}{P_{\mathrm{a}}^{\mathrm{max}}}
\newcommand{\Pbs}{P_{\mathrm{BS}}}
\newcommand{\an}{\alpha_n}
\newcommand{\ans}{\alpha_n^{\star}}

\begin{document}
	
	\title{Spectral Efficiency Optimization for \\Hybrid Relay-Reflecting Intelligent Surface}
	
	\author{\IEEEauthorblockN{Nhan Thanh Nguyen\IEEEauthorrefmark{1},
			Quang-Doanh Vu\IEEEauthorrefmark{2},
			Kyungchun Lee\IEEEauthorrefmark{3}, and
			Markku Juntti\IEEEauthorrefmark{1}}
		\IEEEauthorblockA{\IEEEauthorrefmark{1}Centre for Wireless Communications, University of Oulu, P.O.Box 4500, FI-90014, Finland}
		\IEEEauthorblockA{\IEEEauthorrefmark{2}The Mobile Networks, Nokia,
			90650 Oulu, Finland}
		\IEEEauthorblockA{\IEEEauthorrefmark{3}Dept. of Electrical and Information Engineering, Seoul National University of Science and Technology, Seoul, South Korea}
		(E-mail: nhan.nguyen@oulu.fi, quang-doanh.vu@nokia.com, kclee@seoultech.ac.kr, markku.juntti@oulu.fi)
		\thanks{The research has been supported in part by Academy of Finland under 6Genesis Flagship (grant 318927) and EERA Project (grant 332362). 
			Corresponding author: N. T. Nguyen (email: nhan.nguyen@oulu.fi).}}

	\maketitle

	\begin{abstract}
		We propose a novel concept of hybrid relay-reflecting intelligent surface (HR-RIS), in which a single or few elements are deployed with power amplifiers (PAs) to serve as active relays, while the remaining elements only reflect the incident signals. The design and optimization of the HR-RIS is formulated in a spectral efficiency (SE) maximization problem, which is efficiently solved by the alternating optimization (AO) method. The simulation results show that a significant improvement in the SE can be attained by the proposed HR-RIS, even with a limited power budget, with respect to the conventional reconfigurable intelligent surface (RIS). In particular, the favorable design and deployment of the HR-RIS are analytically derived and numerically justified.
	\end{abstract}
	
	\begin{IEEEkeywords}
    Hybrid relay-reflecting intelligent surface (HR-RIS), Reconfigurable intelligent surface (RIS), MIMO
    \end{IEEEkeywords}

	\section{Introduction}

	A new technology called \emph{reconfigurable intelligent surface} (RIS) has recently emerged as a promising solution to significantly enhance the performance of wireless communication systems \cite{di_renzo_smart_2019}. RIS is often assumed to be designed as a planar meta-surface or realized as a planar array of numerous \emph{passive reflecting elements}, which are connected to a controller, allowing modifying the phases of the incident signals independently in real-time. 
	
	The initial studies on RIS in the literature mainly focus on the performance and design aspects of the RIS-assisted communication systems. Specifically, in \cite{jung2020asymptotic}, the asymptotic achievable rate in a RIS-assisted downlink system is analyzed, and a passive beamformer is proposed to increase the achievable system sum-rate. The gain of RIS is further investigated in \cite{wu2019intelligent} and \cite{wang2020intelligent} for multiple-input-single-output (MISO) downlink channels operated at sub-6 GHz and mmWave frequencies, respectively. It is shown that the RIS of $N$ elements can achieve a total beamforming gain of $N^2$ \cite{wu2019intelligent} and allows the received signal power to increase quadratically with $N$ \cite{wang2020intelligent}. Moreover, the achievable data rate of communication systems assisted by a practical RIS with limited phase shifts and hardware impairments are characterized in \cite{zhang2020reconfigurable, guo2019weighted}.
	
	Another important line of studies on RIS focuses on the capacity/data rate maximization of various RIS-assisted communication systems \cite{gong2020towards}. Specifically, in \cite{yang2020intelligent}, the problem of maximizing the achievable rate of a RIS-enhanced single-input-single-output (SISO)  orthogonal frequency division multiplexing (OFDM) system is solved by jointly optimizing the transmit power allocation and the RIS passive array reflection coefficients. By contrast, the MISO systems are considered in \cite{yang2019irs, yuan2020intelligent, han2019large, di2020practical}. While \cite{yang2019irs, yuan2020intelligent} focus on the joint transmit and passive beamforming problem of RISs with continuous phase shifts to attain substantially increased capacity/data rate, Han \emph{et. al} in \cite{han2019large} consider the discrete phase shifts and justify that only two-bit quantizer is sufficient to guarantee a high capacity, which agrees with the finding in \cite{di2020practical}. Unlike the aforementioned works, the studies in \cite{zhang2020capacity, perovic2021achievable, xiong2020reconfigurable} considered RIS-aided MIMO systems. Specifically, Zhang \emph{et. al} in \cite{zhang2020capacity} develop an efficient alternating optimization (AO) method to find the locally optimal phase shifts of the RIS, providing a significant improvement in the channel total power, rank, condition number, and capacity. Furthermore, Perović \emph{et. al} in \cite{perovic2021achievable} propose the projected gradient method (PGM), which can achieve the same achievable rate as the AO scheme, but with a lower computational complexity. In both \cite{zhang2020capacity, perovic2021achievable}, the perfect channel state information (CSI) is assumed. In contrast, only partial CSI is required in the investigation in \cite{xiong2020reconfigurable}.
	
	A main limitation of the RIS compared to the relays is the fact that the reflection limits the degrees of freedom in the beamforming. Both Bj\"ornson \emph{et. al} \cite{bjornson_intelligent_2019} and Wu \emph{et. al} \cite{wu2019intelligent} show that a very large RIS is needed to outperform decode-and-forward (DF) relaying; otherwise, it can be easily outperformed even by a half-duplex (HD) relay with few elements. Furthermore, as the RIS becomes sufficiently large, the performance improvement is not significant if the number of elements in the RIS increases. This implies that if a few passive elements of the RIS are replaced by active ones, the reduction in the passive beamforming gain is just marginal, while the gain from active relaying can be significant. Furthermore, the introduction of active elements with RF chains and baseband signal processing facilitates efficient channel estimation, as proposed in \cite{taha2019deep, taha2019enabling}. The aforementioned aspects motivate a new architecture of the intelligent meta-surface, allowing it to leverage the benefits of both the RIS and relay, as the HR-RIS proposed in this paper. 
	
	The major contributions of the paper are summarized as follows: We propose the novel HR-RIS architecture, which enables a hybrid active-passive beamforming scheme rather than fully-passive beamforming of the conventional RIS. In particular, HR-RIS exploits the benefits of relaying while mitigating the limitation of passive reflecting. The design and optimization of the HR-RIS is formulated in the SE maximization problem, whose solution is efficiently solved by the AO method. Our analytical results show that this proposed HR-RIS requires activating only a single or few elements of the RIS to serve as active relays to attain remarkable SE improvement. The findings are numerically justified by simulations. The results provide further insights on the fundamental trade-off between RIS and relaying.

	\section{HR-RIS Architecture, System model and SE Problem formulation}
	\label{sec_system_model}
	
	\subsection{Proposed HR-RIS Beamforming Architecture}
	
	The HR-RIS is equipped with $N$ elements, including $M$ passive reflecting and $K$ active relay elements $(M+K=N)$. Here, the passive elements can only shift the phase while the active ones can tune both the phase and amplitude of the incident signal. It is noted that an active element would consume more power for processing compared to the passive one due to the operation of the RF chain. Thus, for practical deployment, we are interested in the case that $1 \leq K  \ll M$. Furthermore, we consider the case that the number and positions of the active elements are predefined and fixed in manufacture. The extension to optimize a HR-RIS with dynamic active elements will be considered in our future work.
	
	Let us denote by $\setA$, $\setA \subset \{1,2,\ldots,N\}$, the set of the positions of the $K$ active elements. Let $\an$ denote the relay/reflection-coefficient used at the $n$th element,
	\begin{align}
	\label{def_alpha}
	\an =
	\begin{cases}
	\abs{\an} e^{j \theta_n}, & \text{if } n \in \setA \\
	e^{j \theta_n}, & \text{otherwise}
	\end{cases},
	\end{align}
	where $\theta_n \in [0, 2\pi)$ represent the phase shift. We note that $\abs{\an} = 1$ for $n \notin \setA$. We also define three diagonal matrices based on $\{\an\}$ as  $\bUpsilon = \text{diag} \{ \alpha_1, \ldots, \alpha_{N} \}$,  $\bPhi = \text{diag} \{ \phi_1, \ldots, \phi_{N} \}$,	and $\bPsi = \text{diag} \{ \psi_1, \ldots, \psi_{N} \}$ where
	\begin{align*}
	\phi_n =
	\begin{cases}
	0, & \text{if } n \in \setA \\
	\an, & \text{otherwise}
	\end{cases}, \text{and } \psi_n =
	\begin{cases}
	\an, & \text{if } n \in \setA \\
	0, & \text{otherwise}
	\end{cases}.
	\end{align*}
	In words, $\bPhi$ and $\bPsi$ contain only the passive reflecting and active relaying coefficients, respectively, while $\bUpsilon$ contains the coefficient of all elements of HR-RIS. Clearly, $\bUpsilon = \bPhi + \bPsi$.
	
	\subsection{System Model}
	\begin{figure}[t]
		\belowcaptionskip = -15pt
		\centering
		\includegraphics[scale=0.55]{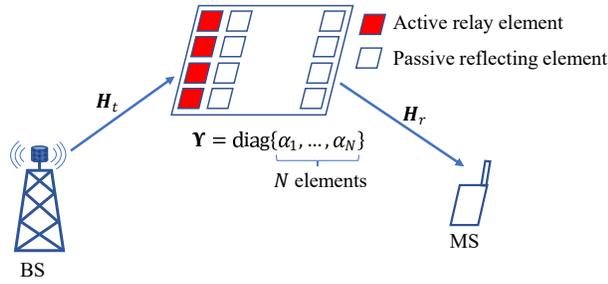}
		\caption{Illustration of the HR-RIS-aided communication system.}
		\label{Fig_HRRIS_system}
	\end{figure}
	We consider downlink transmission where the base station (BS) and mobile station (MS) are equipped with $N_t$ and $N_r$ antennas, respectively, with $N_t, N_r \geq 1$. The communications is aided by the HR-RIS, as illustrated in Fig.\ \ref{Fig_HRRIS_system}. Specifically, the signals reflected more than once on the HR-RIS are assumed to be of negligible power due to the severe pathloss, and, thus, can be ignored \cite{han2020cooperative}. In this work, to focus on the optimization of the semi-passive beamforming at the HR-RIS and to simplify the system model and algorithm design, we assume that the signals propagated on the direct link from the BS to the MS are blocked by obstacles (e.g., buildings, trees) \cite{huang2018achievable, Huang2020Reconfigurable}. Furthermore, we assume the uniform power allocation at the BS. Specifically, let $\vx \in \setC^{N_t \times 1}$ be the transmitted signal vector. Then, we have $\mean{\vx \vx^H} = \Pbs \mI_{N_t}$, where $\Pbs$ is the transmit power at the BS. The signals received by the MS can be given as
	\begin{align*}
	\vy &= \mHr \bPhi \mHt \vx + \mHr \bPsi \mHt \vx + \mHr \bPsi \vn_{\mathrm{H}} + \vn_{\mathrm{MS}}\\
	&= \left(\mHr \bPhi \mHt  + \mHr \bPsi \mHt\right) \vx + \vn \\
	&= \mHr \bUpsilon \mHt \vx  + \vn, \nbthis \label{eq_system_RIS}
	\end{align*}
	where $\mHt \in \setC^{N \times N_t}$ and $\mHr \in \setC^{N_r \times N}$ represents the channels from the BS to the HR-RIS and that from the HR-RIS to the MS, respectively. Furthermore, $\vn_{\mathrm{H}} \sim \mathcal{CN} (0, \sigma_{\mathrm{H}}^2 \mI_{K})$ and $\vn_{\mathrm{MS}} \sim \mathcal{CN} (0, \sigma_{\mathrm{MS}}^2 \mIr)$ are the complex additive white Gaussian noise (AWGN) vectors caused by the $K$ active relay elements at the HR-RIS and $N_r$ receive antennas at the MS, respectively, with $\sigma_{\mathrm{H}}^2$ and $\sigma_{\mathrm{MS}}^2$ being the corresponding average noise variances, and $\vn = \mHr \bPsi \vn_{\mathrm{H}} + \vn_{\mathrm{MS}}$ represents the total effective noise vector at the MS. For simplicity, we assume the same noise power spectrum density and noise figure at the HR-RIS and the MS, yielding $\sigma_{\mathrm{H}}^2 = \sigma_{\mathrm{MS}}^2 \triangleq \sigma^2$ and $\vn \sim \mathcal{CN} \left(0, \sigma^2\left(\mIr + \mHr \bPsi \bPsi^H \mHr^H\right)\right)$.

	\subsection{Problem Formulation}
	
	In this work, the HR-RIS is designed and optimized in terms of SE. Based on \eqref{eq_system_RIS}, the SE can be expressed as
	\begin{align*}
	f_0(\{\alpha_n\}) =  \log_2 \abs{ \mIr + \rho \mHr \bUpsilon \mHt \mHt^H \bUpsilon^H \mHr^H \mR^{-1}},
	\end{align*}
	where $\abs{\cdot}$ denotes the determinant of a matrix or the absolute value of a scalar. $\mR = \mIr + \mHr \bPsi \bPsi^H \mHr^H  \in \setC^{N_r \times N_r}$ is the aggregate noise covariance matrix, and $\rho = \frac{\Pbs}{\sigma^2}$. The transmit power of the active elements of the HR-RIS is  
	\begin{align*}
	\Pa &\triangleq \tr{ \bPsi \left( \mHt \mean{\vx \vx^H} \mHt^H + \sigma^2 \mI_{N} \right) \bPsi^H} \\
	&= \tr{ \bPsi \left( \Pbs \mHt \mHt^H + \sigma^2 \mI_{N} \right) \bPsi^H}. \nbthis \label{eq_Pactive}
	\end{align*}
	The problem of designing coefficients of the HR-RIS for maximizing spectral efficiency can be formulated as
	\begin{subequations}
		\begin{align}
		(\mathrm{P0}) \quad \max_{\{\alpha_n\}} \quad &  f_0(\{\alpha_n\}) \label{opt_hyb_obj_SE} \\
		\textrm{s.t.} \quad 
		& \abs{\alpha_n}=1 \text{ for } n \notin \setA \label{opt_hyb_con_unit_modul}\\
		&\Pa \leq \Pa^{\textrm{max}} \label{opt_hyb_con_relay_power}
		\end{align}
	\end{subequations}
	where $\Pa^{\textrm{max}}$ is the  power budget of the active elements. Function $f_0(\{\alpha_n\})$ is nonconvex with respects to $\{\alpha_n\}$. In addition, the feasible set of $(\mathrm{P0})$ is nonconvex due to the unit-modulus constraint \eqref{opt_hyb_con_unit_modul}. Consequently, problem $(\mathrm{P0})$ is intractable, and it is difficult to find an optimal solution. In the following section, we develop an efficient solution to $(\mathrm{P0})$.
	
	\section{Optimization of the HR-RIS}
	\label{sec_HRRIS}

	To make $(\mathrm{P0})$ tractable, we consider the upper bound of its objective function. Specifically, we have
	\begin{align}
	f_0(\{\alpha_n\})
	&=  \log_2 \abs{ \mIr + \rho \mHr \bUpsilon \mHt \mHt^H \bUpsilon^H \mHt^H \mR^{-1}}  \label{eq_ub_1} \\
	&= \log_2 \abs{ \mR + \rho \mHr \bUpsilon \mHt \mHt^H \bUpsilon^H \mHt^H } - \log_2 \abs{\mR}  \label{eq_ub_2} \\
	&\leq f(\{\alpha_n\})  \label{eq_ub_3},
	\end{align}
	where in \eqref{eq_ub_3}, $f(\{\alpha_n\}) \triangleq \log_2 |\mIr + \mHr \bPsi \bPsi^H \mHr^H  + \rho \mHr \bUpsilon \mHt \mHt^H \bUpsilon^H \mHt^H|$ is obtained by substituting the expression of $\mR$ to the first term in (\ref{eq_ub_2}). The equality occurs when $\setA = \emptyset$. We note that $\log_2 \abs{\mR} = \log_2 \abs{\mI_{N} + \bPsi^H \mHr^H \mHr \bPsi }$ is negligible. The reason is that the path loss is large, and $\bPsi$ is very sparse because HR-RISs are equipped with small numbers of active elements (i.e., small $K$) to ensure a low hardware cost and power consumption for active processing. Thus, $f(\{\alpha_n\})$ is a tight upper bound of $f_0(\{\alpha_n\})$. Therefore, we consider in the sequel the problem of maximizing $f(\{\alpha_n\})$:
	\begin{align*}
	\prob \quad \max_{\{\alpha_n\}} \quad  f(\{\alpha_n\}), \quad
	\textrm{s.t.} \quad \eqref{opt_hyb_con_unit_modul}, \eqref{opt_hyb_con_relay_power}.
	\end{align*}
	
	
	Let $\vr_n \in \setC^{N_r \times 1}$ be the $n$th column of $\mHr$, and let $\vt_n^H \in \setC^{1 \times N_t}$ be the $n$th row of $\mHt$. Then, we have $\mHr = [\vr_1, \ldots, \vr_N]$ and $\mHt = [\vt_1, \ldots, \vt_N]^H$. Because of the diagonal structure of $\bUpsilon$ and $\bPsi$, we have $\mHr \bUpsilon \mHt = \sum_{n=1}^{N} \an \vr_n \vt_n^H$ and $\mHr \bPsi = \sum_{n \in \setA} \an \vr_n$, which leads to
	\begin{align*}
	&\mIr + (\mHr \bPsi) (\mHr \bPsi)^H +  \rho(\mHr \bUpsilon \mHt) (\mHr \bUpsilon \mHt)^H  \\
	&= \mA_n +  \abs{\an}^2 \mB_n + \an \mC_n + \an^* \mC_n^H, \nbthis \label{eq_ABC}
	\end{align*}
	where $\mA_n$, $\mB_n$, and $\mC_n$ are defined as follows:
	\begin{align*}
	\mA_n &\triangleq \mIr + \left( \sum_{i \in \setA, i \neq n} \alpha_i \vr_i \right) \left( \sum_{i\in \setA,i \neq n} \alpha_i^* \vr_i^H\right)  \\
	&\hspace{1cm}+ \rho \left( \sum_{i=1, i \neq n}^{N} \alpha_i \vr_i \vt_i^H\right) \left( \sum_{i=1,i \neq n}^{N} \alpha_i^* \vt_i \vr_i^H\right), \nbthis \label{def_A_active} \\
	\mB_n &\triangleq \vr_n \vr_n^H + \rho \vr_n \vt_n^H \vt_n \vr_n^H , \nbthis \label{def_B_active} \\
	\mC_n &\triangleq \vr_n  \left( \sum_{i\in \setA, i \neq n} \alpha_i^* \vr_i^H \right) + \rho \vr_n  \vt_n^H \left( \sum_{i=1, i \neq n}^{N} \alpha_i^*  \vt_i \vr_i^H\right)  \nbthis \label{def_C_active} 
	\end{align*}
	for $n \in \setA$, and
	\begin{align*}
	\mA_n &\triangleq \mIr + \rho \left( \sum_{i=1, i \neq n}^{N} \alpha_i \vr_i \vt_i^H\right) \left( \sum_{i=1,i \neq n}^{N} \alpha_i^* \vt_i \vr_i^H\right), \nbthis \label{def_A_passive}  \\
	\mB_n &\triangleq \rho \vr_n \vt_n^H \vt_n \vr_n^H ,  \nbthis \label{def_B_passive}\\
	\mC_n &\triangleq \rho \vr_n \vt_n^H \left( \sum_{i=1, i \neq n}^{N} \alpha_i^* \vt_i \vr_i^H\right) \nbthis \label{def_C_passive}
	\end{align*}
	for $n \notin \setA$. From \eqref{eq_ABC}, the objective function of $\prob$ can be rewritten in the form
	\begin{align}
	f(\alpha_n) \triangleq \log_2 \abs{\mA_n +  \abs{\an}^2 \mB_n + \an \mC_n + \an^* \mC_n^H}, \label{fm}
	\end{align}
	where matrices $\mA_n, \mB_n,$ and $\mC_n$ are all independent of one particular $\an$, although they depend on other coefficients $\alpha_i, i \neq n$. It is observed that the new objective function $f(\alpha_n)$ of $\prob$ is in an explicit form of \emph{an individual reflecting/relaying coefficient} $\an, n=1,\ldots,N$. This motivates an AO method  \cite{zhang2020capacity} to solve $\prob$ by optimizing each values of $\an, n=1,\ldots,N$\footnote{We note that the coefficients $\{\an\}$ can also be obtained by employing the PGM scheme \cite{perovic2021achievable}. However, we adopt the AO method because it can provide closed-form solutions to $\{\an\}$, which facilitates the optimal structure design of the HR-RIS.}. Furthermore, because of the diagonal structure of $\bPsi$, we have $\tr{\bPsi \bPsi^H} = \sum_{n \in \setA} \abs{\an}^2$ and $\tr{\bPsi \mHt \mHt^H \bPsi^H} = \sum_{n \in \setA} \abs{\an}^2 \norm{\vt_n}^2$. Therefore, after some manipulations, $\Pa$ in \eqref{eq_Pactive} can be further expanded as
	\begin{align*}
	\Pa = \sum_{n \in \setA} \abs{\an}^2 \xi_n, \nbthis \label{eq_relay_power}
	\end{align*}
	where $\xi_n \triangleq \sigma^2 + \Pbs \norm{\vt_n}^2$.	From \eqref{fm} and \eqref{eq_relay_power}, $\an$ is the solution to the problem
	\begin{subequations}
		\begin{align*}
		\probn \quad \max_{\an} \quad & f(\alpha_n) \nbthis \label{opt_f_obj} \\
		\textrm{s.t.} \quad & \abs{\an} = 1, n \notin \setA \nbthis \label{opt_f_con_passive}\\
		& \abs{\an}^2 \leq \frac{\Pamax - \tilde{\Pa}}{\xi_n}, n \in \setA. \nbthis \label{opt_f_con_active}
		\end{align*}
	\end{subequations}
	where $\tilde{\Pa} \triangleq \sum_{i \in \setA, i \neq n} \abs{\alpha_i}^2 \xi_i$ represents the total power allocated to the other active elements, which is a constant because $\{\alpha_i\}_{i \in \setA, i \neq n}$ are fixed in the considered AO method. This optimization problem can be further simplified by noting that $\mA_n$ is a full rank matrix $(\textrm{rank}(\mA_n) = N_r)$ and invertible based on its definition. Therefore, $f(\alpha_n)$ can be expressed as
	\begin{align*}
	&f(\alpha_n)  = \log_2 \abs{\mA_n} + g(\alpha_n), \nbthis \label{def_fn}
	\end{align*}
	where $g(\alpha_n) \triangleq \log_2 |\mIr +  \abs{\an}^2 \mA_n^{-1}\mB_n + \an \mA_n^{-1}\mC_n + \an^* \mA_n^{-1}\mC_n^H|$. Because $\log_2 \abs{\mA_n}$ is independent of $\an$, the problem in \eqref{opt_f_obj} is equivalent to $\max_{\an}  g(\alpha_n), \textrm{s.t.}  \eqref{opt_f_con_passive}, \eqref{opt_f_con_active}$. Therefore, we focus on investigating $g(\alpha_n)$ in the sequel. 
	
	Let $\mD_n = \mIr + \abs{\an}^2 \mA_n^{-1}\mB_n$. Then, we have
	\begin{align*}
	g(\alpha_n) &= \log_2 \abs{\mD_n + \an \mA_n^{-1}\mC_n + \an^* \mA_n^{-1}\mC_n^H}\\
	&= \log_2 \abs{\mD_n} + h(\alpha_n) , \nbthis \label{def_h_n}
	\end{align*}
	where $h(\alpha_n) \triangleq \log_2 \abs{ \mIr + \an \mE_n^{-1}\mC_n + \an^* \mE_n^{-1}\mC_n^H}$ with $\mE_n \triangleq \mA_n \mD_n$. We at first investigate the first term in \eqref{def_h_n}, i.e., $\log_2 \abs{\mD_n} = \log_2 \abs{\mIr + \abs{\an}^2 \mA_n^{-1}\mB_n}$. We note that $\mathrm{rank} (\mA_n^{-1}\mB_n) \leq \mathrm{rank} (\mB_n) = 1$. Therefore, $\mA_n^{-1}\mB_n$ has rank of either zero or one. Because $\mathrm{rank} (\mA_n^{-1}\mB_n) = 0$ only when $\mA_n^{-1}\mB_n = \textbf{0}$, we focus on the case $\mathrm{rank} (\mA_n^{-1}\mB_n) = 1$. Furthermore, $\mA_n^{-1}\mB_n$ is non-diagonalizable if and only if $\mathrm{trace} (\mA_n^{-1}\mB_n) = 0$ \cite{zhang2020capacity}, which rarely happens in general. Therefore, we focus on the case that $\mathrm{trace} (\mA_n^{-1}\mB_n) \neq 0$ and $\mA_n^{-1}\mB_n$ is diagonalizable. As a result, $\mA_n^{-1}\mB_n$ can be expressed as $\mA_n^{-1}\mB_n = \mQ_n \boldsymbol{\Gamma}_n \mQ_n^{-1}$ based on the eigenvalue decomposition (EVD), with $\mQ_n \in \setC^{N_r \times N_r}$ and $\boldsymbol{\Gamma}_n = \diag{\gamma_n, 0, \ldots, 0}$. Here, $\gamma_n$ is the sole non-zero eigenvalue of $\mA_n^{-1}\mB_n$. Furthermore, because $\mA_n$ and $\mB_n$ are both positive semi-definite, $\gamma_n$ is non-negative and real, and we obtain
	\begin{align*}
	\log_2 \abs{\mD_n} &= \log_2 \abs{\mIr + \abs{\an}^2 \mQ_n \boldsymbol{\Gamma_n} \mQ_n^{-1}} \\
	&= \log_2 \left(1 + \abs{\an}^2  \gamma_n \right). \nbthis \label{eq_det_Dn}
	\end{align*}

	Considering the second term in \eqref{def_h_n}, i.e., $h(\alpha_n)$, we also focus on the case that $\mE_n^{-1}\mC_n$ is diagonalizable, which allows it to be decomposed as $\mE_n^{-1}\mC_n = \mU_n \boldsymbol{\Lambda}_n \mU_n^{-1}$ based on the EVD, with $\mU_n \in \setC^{N_r \times N_r}$, $\boldsymbol{\Lambda}_n = \diag{\lambda_n, 0, \ldots, 0}$, and $\lambda_n$ is the sole non-zero eigenvalue of $\mE_n^{-1}\mC_n$. Furthermore, let $\mV_n \triangleq \mU_n^H \mE_n \mU_n$, $v_n$ be the first element of the first column of $\mV_n^{-1}$, and $v_n^{\prime}$ be the first element of the first row of $\mV_n$. Then, $h(\alpha_n)$ can be expressed as \cite{zhang2020capacity}:
	\begin{align}
	h(\alpha_n) = \log_2 \left(1 + \abs{\an}^2 \abs{\lambda_n}^2  + 2\re{\an \lambda_n} - v_n^{\prime} v_n \abs{\lambda_n}^2 \right), \label{eq_g_x}
	\end{align}
	where $\re{\cdot}$ denotes the real part of a complex number. The detailed derivation of \eqref{eq_g_x} can be found in \cite{zhang2020capacity}. We note that, compared to \cite{zhang2020capacity}, the additional coefficient $\abs{\an}^2$ is due to the active relaying coefficient in the HR-RIS, which does not exist for the conventional RIS  \cite{zhang2020capacity}. 
	
	From \eqref{def_fn}--\eqref{eq_g_x}, we obtain
	\begin{align*}
	&f(\alpha_n) = \log_2 \abs{\mA_n} + \log_2 \left(1 + \abs{\an}^2  \gamma_n \right) \\
	&+ \log_2 \left(1 + \abs{\an}^2 \abs{\lambda_n}^2  + 2\re{\an \lambda_n} - v_n^{\prime} v_n \abs{\lambda_n}^2 \right). \nbthis \label{eq_fn}
	\end{align*}
	It is observed from \eqref{eq_fn} that the phase of the optimal solution $\ans$ to $\probn$ is  $-\arg \{ \lambda_n \}$, and $f(\alpha_n)$ is a monotonically increasing function of $\abs{\ans}$. Therefore, the optimal value of $\abs{\ans}$ is given as $\abs{\ans} = \frac{\Pamax - \tilde{\Pa}}{\xi_n}, \forall n \in \setA$ based on \eqref{opt_f_con_active}. Hence, the optimal coefficients of the HR-RIS is given by
	\begin{align}
	\label{eq_alpha_best_fix}
	\ans =
	\begin{cases}
	\sqrt{\frac{\Pamax - \tilde{\Pa}}{\xi_n}} e^{-j \arg \{ \lambda_n \}}, & n \in \setA\\
	e^{-j \arg \{ \lambda_n \}}. & \text{otherwise}
	\end{cases},
	\end{align}
	allowing $\bUpsilon^{\star}$ to be solved as in Algorithm \ref{alg_fix}. Specifically, in step 1, the coefficient matrix $\bUpsilon$ is initialized with random phases, and random power fractions are allocated to the active elements. As a result, in steps 2--10, the coefficients of the HR-RIS are updated based on \eqref{eq_alpha_best_fix} until the objective value is converged.  Finally,  $\bUpsilon^{\star}$ is obtained in step 11. The monotonic convergence of Algorithm \ref{alg_fix} is guaranteed because the objective value of $\probn$ is non-decreasing over iterations \cite{zhang2020capacity}. 
	
	\begin{remark}
		\label{remark}
		The result in \eqref{eq_alpha_best_fix} suggests the favorable deployment of the HR-RIS. Specifically, the power amplifier coefficient $\abs{\an}$ of an active element decreases with $\tilde{\Pa}$ and $\xi_n$. Therefore, with a limited power budget $\Pamax$, the HR-RIS with a small number of elements, i.e., small $K$, is easier to attain SE gains than that with numerous active elements. Furthermore, with a fixed $\Pamax$, a lower $\Pbs$ and/or smaller channel gain $\norm{\vt_n}^2$ from the BS, which can be caused by a more severe pathloss, results in a higher power amplifier coefficients $\abs{\ans}$, and thus, more significant performance improvement.
	\end{remark} 
	
	\begin{algorithm}[t]
		\caption{Find $\bUpsilon^{\star}$ for HR-RIS}
		\label{alg_fix}
		\begin{algorithmic}[1]
			\REQUIRE $\mHt, \mHr, \setA$.
			\ENSURE $\bUpsilon^{\star}$.
			\STATE Randomly generate $\{\alpha_n\}$ with $\abs{\an}=1$ for $n \notin \setA$ and $\sum_{n \in \setA} \abs{\alpha_n}^2 \xi_n = \Pamax$ for $n \in \setA$.
			\WHILE{not converge}
			\FOR {$n = 1 \rightarrow N$}
			\STATE Compute $\mA_n, \mB_n$, and $\mC_n$ based on \eqref{def_A_active}-\eqref{def_C_passive}.
			\STATE $\mD_n = \mIr + \abs{\an}^2 \mA_n^{-1}\mB_n$, $\mE_n = \mA_n \mD_n$.
			\STATE Find $\lambda_n$ as the sole non-zero eigenvalue of $\mE_n^{-1}\mC_n$.\\
			\STATE Compute $\ans$ based on \eqref{eq_alpha_best_fix}.
			\ENDFOR 
			
			\STATE Check convergence.
			\ENDWHILE
			\STATE $\bUpsilon^{\star} = \diag{\alpha_1^{\star}, \ldots, \alpha_N^{\star}}$.
		\end{algorithmic}
	\end{algorithm}


	\section{Simulation Results}
	\label{sec_sim}
	\begin{figure}[t]
		\centering
		\includegraphics[scale=0.6]{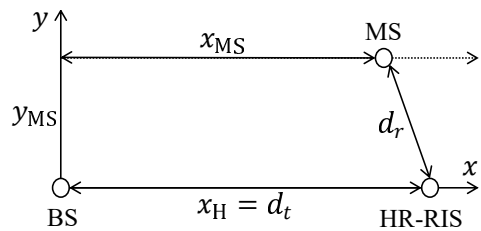}
		\caption{Horizontal locations of the BS, MS, and HR-RIS.}
		\label{fig_system_illustration}
	\end{figure}
	
	In this section, numerical results are provided to validate the proposed HR-RIS scheme. We assume that in a two-dimensional coordinate, the BS, HR-RIS, and MS are deployed at $(0,0)$, $(x_{\mathrm{H}},0)$, and $(x_{\mathrm{MS}},y_{\mathrm{MS}})$, respectively, as illustrated in Fig.\ \ref{fig_system_illustration}. As a result, the distance between the BS and MS is $d_t = x_{\mathrm{H}}$, and the corresponding path loss is given by $\beta(d_t) = \beta_0 \left(\frac{d_t}{1\mathrm{ m}}\right)^{\epsilon_t}$, where $\beta_0 = -30$ dB is assumed \cite{wu2019intelligent, zhang2020capacity}, and $\epsilon_t$ is the path loss exponent. The small scale-fading channel between the BS and the HR-RIS is modeled by $\tilde{\mH}_t = \left(\sqrt{\frac{\kappa_t}{1+\kappa_t}} \mHt^{\mathrm{LoS}} + \sqrt{\frac{1}{1+\kappa_t}} \mHt^{\mathrm{NLoS}}\right)$ following the Rician fading channel model \cite{wu2019intelligent, zhang2020capacity}, where $\mHt^{\mathrm{LoS}}$ and $\mHt^{\mathrm{NLoS}}$ represent the deterministic LoS and NLoS components, respectively. The channel matrix between the BS and the HR-RIS, i.e., $\mHt$, is obtained by multiplying $\tilde{\mH}_t$ by the square root of the corresponding path loss $\beta(d_t)$, i.e., $\mHt = \sqrt{\beta(d_t)} \tilde{\mH}_t$. The channel matrix between the HR-RIS and MS is modeled similarly. For the detailed generation of $\mHt^{\mathrm{LoS}}$ and $\mHt^{\mathrm{NLoS}}$, please refer to \cite{wu2019intelligent, zhang2020capacity}. 
	
	In simulations, we set $\{\epsilon_t,\epsilon_r\} = \{2.2,2.8\}$ and $\{\kappa_t,\kappa_r\} = \{\infty,0\}$ \cite{wu2019intelligent}, where the subscripts $(\cdot)_t$ and $(\cdot)_r$ imply the parameters associated with $\mHt$ and $\mHr$, respectively. Furthermore, $\sigma^2 = -80$ dBm, $\{x_{\mathrm{MS}}, y_{\mathrm{MS}}, x_{\mathrm{H}} \} = \{40,2,51\}$ m, and two-bit resolution phase shifts of the RIS/HR-RIS are assumed unless otherwise stated. For comparison, we consider the conventional RIS with $N$ fully passive reflecting elements, whose phases are either randomly generated or optimized using the AO method in \cite{zhang2020capacity}.
	
	\begin{figure*}[t]
		\centering
		\includegraphics[scale=0.66]{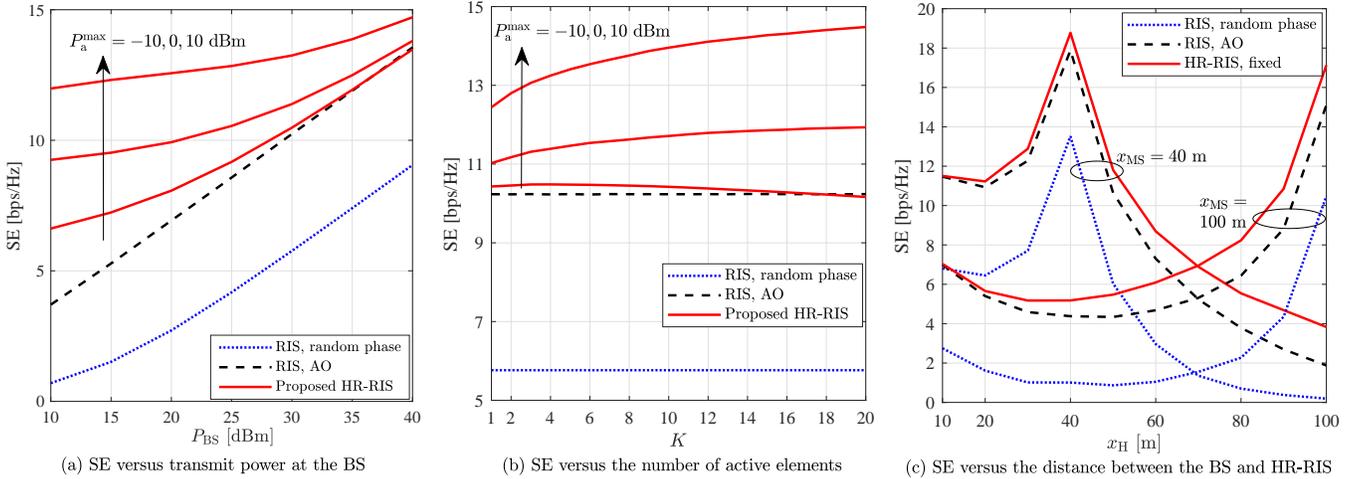}
		\caption{SEs of the proposed HR-RIS scheme compared to those of the conventional RIS schemes for $N_t = 32$, $N_r = 2$, $N = 50$. The other parameters are set as follows: $K = 4$ in Fig.\ 3(a); $\Pbs = 30$ dBm in Fig.\ 3(b); $K = 4$, $\Pamax = 0$ dBm, and $\Pbs = 30$ dBm in Fig.\ 3(c).}
		\label{fig_rate}
	\end{figure*}

	In Fig.\ \ref{fig_rate}(a), the SEs of a $32 \times 2$ MIMO system with $N=50$, $K = 4$, $M=N-K=46$, and $\Pamax=\{-10, 0, 10\}$ dBm is shown. It is observed that the RIS with random phases has poor performance compared to the optimized RIS. By contrast, the HR-RIS scheme achieves significant improvement in the SE with respect to the conventional RIS, especially at low $\Pbs$, agreeing with Remark \ref{remark}. The HR-RIS only requires a small $\Pamax$ to achieves remarkable SE improvement at low $\Pbs$, and with $\Pamax = \{0, 10\}$ dBm, sustainable SE gains of the HR-RIS are seen for the whole considered range of $\Pbs$. 
	
	The SE improvements of the HR-RIS for various $K$ are shown in Fig.\ \ref{fig_rate}(b) with $\Pamax=\{-10, 0, 10\}$ dBm and $\Pbs = 30$ dBm. It is seen that deploying more active elements does not always guarantee a higher SE. Specifically, increasing $K$ improves the SE for $\Pamax=\{0, 10\}$ dBm, which, however, degrades the SE for $\Pamax=-10$ dBm. In general, a satisfying SE performance gain with respect to the RIS can be attained even with small $K$, or even with $K=1$.
	
	In Fig.\ \ref{fig_rate}(c), we present the SEs of the HR-RIS for different geographical deployment: the MS is at $\{(40,2), (100,2)\}$, while the HR-RIS/RIS is located at $(0, x_{\mathrm{H}})$, where $x_{\mathrm{H}} \in [10, 100]$ m. Increasing $x_{\mathrm{H}}$, i.e., the RIS/HR-RIS moves far away from the BS, causing more severe pathloss between the BS and HR-RIS, and hence, more significant active beamforming gains are attained, as discussed in Remark \ref{remark}. It can be concluded from Fig.\ \ref{fig_rate}(c) that the proposed HR-RIS is robust over geographical deployment.

	\section{Conclusion}
	\label{sec_concusion}
	
	We proposed a novel HR-RIS to assist MIMO communication systems with significantly improved SE compared to that assisted by the conventional RIS. The HR-RIS is a semi-passive beamforming architecture, in which a few elements are capable of adjusting the incident signals' power. It has been designed and optimized by the AO scheme. The analytical results suggest the favorable design and deployment of the HR-RIS. Specifically, a small number of active elements is sufficient for a satisfying SE gain with respect to the conventional RIS. Furthermore, the HR-RIS can attain significant SE improvement when the transmit power at the transmitter is small or moderate, and the improvement is more significant when the distance between the transmitter and the HR-RIS is large. Finally, intensive simulations have been performed to numerically justify the findings. The results reveal that the HR-RIS has the potential to outperform both the passive RIS and active AF relaying in certain communications channels. Further investigation and optimization of the proposed HR-RIS architecture in terms of energy efficiency as well as the consideration of transmit beamforming at the BS can be considered for future research.
	
	\section*{Acknowledgment}
    This work has been supported in part by Academy of Finland under 6Genesis Flagship (grant 318927) and EERA Project (grant 332362).

	\bibliographystyle{IEEEtran}
	\bibliography{IEEEabrv,Ref}

\end{document}